\documentclass[english, 12pt]{article}
\usepackage[round]{natbib}
\usepackage[T1]{fontenc}
\usepackage[latin9]{inputenc}
\usepackage{geometry, bm}
\usepackage{comment}
\usepackage{mathrsfs}
\usepackage{amsmath}
\usepackage{amssymb}
\usepackage{setspace}
\usepackage{graphicx}
\usepackage{babel}

\makeatletter
\usepackage{babel}

\makeatother

\begin{document}

\title{The Post Double LASSO for Efficiency Analysis\thanks{Helpful comments by the participants of NAPW 2023, EWEPA XVII, ESAM2024 and Machine Learning in Business workshop at the University of Sydney are gratefully acknowledged. Authors acknowledge the financial support from ARC Discovery Grants. Prokhorov acknowledges support from an RNF grant. Zelenyuk acknowledges support from Australian Research Council (grant FT170100401)}
}

\author{Christopher F. Parmeter\thanks{Miami Herbert Business School, University of Miami, USA; e-mail: cparmeter@bus.miami.edu.} \\
 \and Artem Prokhorov\thanks{University of Sydney Business School; email: artem.prokhorov@sydney.edu.au.} \thanks{Center for Econometrics and Business Analytics (CEBA); St.Petersburg State University, Russia.} \thanks{Centre interuniversitaire de recherche en economie quantitative (CIREQ); University of Montreal, Canada.} \\
 \and Valentin Zelenyuk\thanks{School of Economics and Centre for Efficiency and Productivity Analysis, The University of Queensland, Australia; email: v.zelenyuk@uq.edu.au.}\\
}

\maketitle
\vspace{-1cm}

\begin{abstract}

Big data and machine learning methods have become commonplace across economic milieus. One area that has not seen as much attention to these important topics yet is efficiency analysis. We show how the availability of big (wide) data can actually make detection of inefficiency more challenging. We then show how machine learning methods can be leveraged to adequately estimate the primitives of the frontier itself as well as inefficiency using the `post double LASSO' by deriving Neyman orthogonal moment conditions for this problem. Finally, an application is presented to illustrate key differences of the post-double LASSO compared to other approaches.

\end{abstract}
\medskip{}
 \textbf{\footnotesize{}Key Words:}{\footnotesize{} {Shrinkage
and Selection, Neyman Orthogonality, Stochastic Frontier Analysis, Moment Parameter Redundancy.}
\vskip 0.1cm}
\noindent\textbf{\footnotesize{}{}JEL Classification}{\footnotesize{}{}:
{C1,C14, C13}} \newpage\setlength{\textwidth}{15.2 cm} 
\global\long\def\baselinestretch{1.3}%
 \pagestyle{plain} \setcounter{page}{1} \setcounter{footnote}{0}

\onehalfspacing

\section{Introduction}

Evaluating policy (both prescriptive and proscriptive) is a fundamental task of economists. One aspect of any good policy design is the impact that firm level inefficiency has on overall outcomes. To see the importance of this look no further than various empirical literatures studying hospitals, banks, farms, airline travel, energy and water utilities, and distribution networks. Debates continually arise as to how best to regulate these industries, how they should be structured, limiting monopoly behavior and ensuring adequate protections for consumers. The list goes on. 

At the center of such debate is the existence of inefficiencies in the production process. Basic microeconomic theory would suggest that, under certain (and somewhat restrictive) assumptions, inefficiencies are eliminated by the market -- competitive forces work to rid the market of those producers who are not fully optimizing. And yet, such a thought is at variance with basic intuition from casual observation. Consider the recent work of \cite{hanna/etal:14}, who studied seaweed farmers via learning through noticing (similar to rational inattention, \cite{mackowiak/etal:23}) in the context of selective attention. 
What they find is that (pg. 1312) ``A farmer planting a crop faces a slew of potential features\dots He cannot possibly attend to everything \dots his attention is limited while the number of potentially important variables is large. Since he can only learn about the dimensions that he notices (or attends to), this choice becomes a key input into the learning process.''

This selective attention is important to understand from a modeling perspective. Consider that even a farmer with considerable experience may not reach the production frontier. Why? A farmer who initially 
believes that an important dimension of farming is \textit{ex ante} unimportant will not attend to it. Subsequently, they will not be able to learn if this dimension matters and as such they enter a feedback loop where they cannot learn about their false beliefs of production. From a modeling standpoint these dimensions can be easily accounted for in a regression setting, by adding in more `control' (or `conditioning') variables to capture these potentially important features that farmers are overlooking. However, the various dimensions to be accounted for can quickly overwhelm the ability of an estimator to be informative. 

Now, suppose we are concerned both with identifying inefficiency as well as these overlooked dimensions of production process and their conditions or environment. As we will discuss in the next section, when additional variables are added to characterize the production process, this has the very real consequence of practically eliminating the inefficiency signal (relative to noise) in the stochastic frontier (SF) model of \cite{ALS:77}. When it comes to handling this increasing (wide) data size, one might immediately turn to machine learning and predictive analytics, which recently have become quite popular across a range of application domains in economics. However, their widespread use in efficiency analysis has lagged and has only very recently been introduced \citep[see, e.g., ][to name a few]{jin/lee:18, horrace/etal:23, TPZ_2023_JoE}. 

One can hypothesize why this might be the case, but a simple explanation may prove most elucidating. To explain why inefficiency arises it is common to collect and include a variety of explanatory variables that are hypothesized to raise or lower inefficiency (or the frontier to which it is measured): a dummy variable for public vs. private, the age of the manager, if the work force is unionized, the years of education of the manager, dummy variables for the geographic location of the firm, etc. But an unfortunate consequence of inclusion of a large number of these variables is that it makes it much more difficult to detect inefficiency. This may seem paradoxical but it has an intuitive explanation, which we will document more forcefully below. As we include more variables in our analysis (think wide data), there is less room in the error term and parsing out inefficiency is implicitly linked to explaining skewness in the error term. Without this skewness, inefficiency is not identified in the mainstream stochastic frontier paradigm and hence no inefficiency is found there. Even ignoring skewness, the intuition is clear. As more variables are included to explain output, there is less in the residual. As the residual gets closer to zero, so too does inefficiency. But if the variables that are being included influence neither output nor inefficiency, then we miss the ability to learn about inefficiency.



We start by exploring whether inefficiency can arise spuriously as a result of a large number of irrelevant factors present in a simple SF model estimated by corrected ordinary least squares (COLS). We then show that LASSO can effectively detect irrelevant variables and re-establish the true skewness of the error term. However, this fix has its own caveats as there are well-known issues with direct application of the LASSO when one is interested in estimation of a specific parameter value \citep[see, e.g.,][]{belloni:13}. Hence, we explore and discuss the impact that application of the LASSO has on the estimates stemming from maximum likelihood estimation of the stochastic frontier model. Not surprisingly, we see that the same issues in the regression setting with direct application of the LASSO prior to MLE leads to a bias in the estimation of ``fixed'' parameters of the frontier model. This leads us to adapt a post-double LASSO estimator from \cite{belloni:13} for the stochastic frontier model, which helps resolve this bias issue. In particular, in a Generalized Method of Moments (GMM) framework, we derive a set of moment conditions that cannot be improved upon by using either the LASSO moment conditions or true value of the parameters estimated by LASSO (M/P-redundancy). We show how the M/P redundancy is related to the Neyman orthogonality of post-double LASSO, which ensures that the biases induced by direct application of the LASSO are removed. Finally, evidence from an analysis of Milk farming in Spain is presented to illustrate key differences of the post-double LASSO compared to other approaches.




The plan for the paper is as follows. Section 2 discusses the benchmark stochastic frontier model and covers various estimators available. It also contains illustrative simulations to highlight the issue of data width versus the level of inefficiency. Section 3 provides theoretical insights into our post double LASSO estimator for maximum likelihood estimation of the cross-sectional stochastic frontier model. Section 4 contains an illustrative example with real data highlighting the issues discussed in the paper and the proposed solutions. Section 5 concludes. 

\section{Basic Framework}

The basic stochastic frontier model we consider is
\begin{equation}\label{eq:sfa}
  y_i =\beta_0+\bm x_i\beta + \bm z_i\delta+ v_i- u_i=\beta_0+\bm x_i\beta +\bm z_i\delta+ \epsilon_i,
\end{equation}
where $y_i$ is output, $\bm x_i$ is a $p$-dimensional vector of production inputs, $\bm z_i$ is a $d$-dimensional vector that includes other potential influences on production where $d$ may be quite large (certainly $d>>p$) and $\epsilon_i=v_i-u_i$ is the composite error terms composed of a two-sided noise, $v_i\overset{iid}{\sim} N(0, \sigma_v^2)$, and inefficiency, $u_i\overset{iid}{\sim} N_+(0, \sigma_u^2)$, where $i=1, \ldots, n,$ is the indexes of the production units, which we will sometimes omit for brevity. Conventional specifications have a handful of inputs, e.g., three (labor, capital and materials), so the unknown parameter vector $\beta$ contains an intercept and three slope coefficients. 

To help illustrate the issues present when $\bm z$ enters the stochastic frontier model, we discuss two widely popular estimators for $\beta-0$, $\beta$, $\delta$, $\sigma_v$ and $\sigma_u$ which are common in low-dimensional settings.


\subsection{Corrected Ordinary Least Squares}\label{sec:cols}

COLS is a regression of $y$ on $\bm x$ and $\bm z$ \cite[see][]{ALS:77, olson/eta:80}\footnote{Strictly speaking, these papers did not include $\bm z$ when they introduced COLS.} followed by a correction of the intercept, based on moment conditions which depend on the specific distributional assumptions made \cite{PARMETER:2023}. It is known to produce a consistent estimator of the slope coefficients with no distributional assumptions on $v$ and $u$ (apart from being iid with finite variances, zero mean for $v$ and independent from $\bm x$ and $\bm z$) but the intercept estimator is biased due to the nonzero mean of $\epsilon$. That is, the intercept in the regression model which OLS estimates is $\beta_0 - E(\epsilon_i)$ and one needs distributional assumptions to estimate $\beta_0$ by correcting the OLS estimate of the intercept. 

Using the assumption that $v_i$ is Normal, $u_i$ is Half-Normal and $v_i$ and $u_i$ are independent, we can write the first three moments of $\epsilon_i$ as follows
\[\mu_1\equiv E(\epsilon_i) = \sqrt{\frac{2}{\pi}}\sigma_u, \quad \mu_{2}\equiv V(\epsilon_i) = \sigma_v^2 + \frac{\pi-2}{\pi}\sigma_u^2, \quad {\mu}_{3} \equiv E(\epsilon_i-E(\epsilon_i))^3 = \frac{\pi-4}{\pi}\sqrt{\frac{2}{\pi}}\sigma_u^3.\]
Note that the error term is asymmetric with a negative skew in the population, $\mu_3<0$, as long as there is inefficiency in the model, $\sigma_u \ne 0$. So, asymmetry of the error is what reveals the presence of inefficiency. Also note that the three moments depend on two unknown parameters $\sigma_u$ and $\sigma_v$, and an estimate of $\sigma_u$ is required to correct the OLS estimate of the intercept. 

Define the OLS residuals $e_i = y_i - \hat\beta_0-{x}_i\hat{\beta}-z_i\hat{\delta}$, where $(\hat{\beta}, \hat{\delta})$ are OLS estimates of $(\beta, \delta)$. It is easy to see that the population moments above can be estimated using the second- and third-order sample moments of $e_i$. Specifically, $\hat{\mu}_{2} \equiv \frac{1}{n}\sum_{i=1}^n e_i^2$ and $\hat{\mu}_{3} \equiv \frac{1}{n}\sum_{i=1}^n e_i^3$ are consistent estimators for $\mu_2$ and $\mu_3$, respectively. Solving for the unknown parameters in terms of sample quantities, one obtains the COLS estimators of $\sigma_u^2$ and $\sigma_v^2$:
\[\hat{\sigma}^2_u \equiv \left(\frac{\pi}{\pi-4}\sqrt{\frac{\pi}{2}}\hat{\mu}_3\right)^{2/3} \quad \text{and}\quad \hat{\sigma}_v^2\equiv \hat{\mu}_2 -\frac{\pi-2}{\pi}\hat{\sigma}^2_u.\]
The well known wrong skew problem \citep{Waldman:82} arises when $\hat{\mu}_3>0$ even though $\mu_3<0$ in the population, in which case it is common to set $\hat{\sigma}_u^2=0$. With the estimator $\hat{\sigma}_u$, the COLS estimator of $\beta_0$ is the OLS estimator of the intercept, $\hat\beta_0$ plus $\sqrt{\frac{2}{\pi}}\hat{\sigma}_u$. 

\subsection{Maximum Likelihood Estimation}
Maximum likelihood estimation (MLE) is arguably the most widely used estimator in practice when it comes to asymmetric error distributions. It obtains all parameter estimates simultaneously by maximizing the likelihood over $\Theta=(\beta_0,\beta, \delta, \sigma_u^2, \sigma_v^2)$ \citep{ALS:77, meeusen/vandenbroeck:77}:
\begin{equation}\label{MLE}
\underset{\Theta}{\max}\sum\limits^n_{i=1}\ell(\epsilon_i;\Theta),
\end{equation}
where 
\begin{equation*}
    \ell(\epsilon_i;\Theta)=\ln\sigma+\ln\Phi(-\epsilon_i\lambda/\sigma)-\frac{\epsilon_i^2}{2\sigma^2}
\end{equation*}
is the individual log-likelihood for observation $i$, $\epsilon_i=y_i-\beta_0-\bm x_i\beta-\bm z_i\delta$ and we have used the common notation $\sigma^2=\sigma_u^2+\sigma_v^2$ and $\lambda=\sigma_u/\sigma_v$. An alternative parameterization due to \cite{battese/corra:77} uses $\gamma=\sigma^2_u/\sigma^2$ and has individual likelihood
\begin{equation*}
    \ell(\epsilon_i;\Theta)=\ln\sigma+\ln\Phi\left(-\sqrt{\frac{\gamma}{1-\gamma}}\left(\epsilon_i/\sigma\right)\right)-\frac{\epsilon_i^2}{2\sigma^2}.
\end{equation*}

\subsection{No inefficiency with wide data?}

Both COLS and MLE are computationally troublesome when $p$ is large and infeasible when $d+p \ge n$. Moreover, and even more concerning, is that the estimation of $\sigma_u$ breaks down when $d+p$ approaches $n$. We illustrate the effect of high-dimensional covariates $\bm z$ on the asymmetry of $\varepsilon_i$ in the context of COLS, which is equivalent to MLE as per \cite{Waldman:82} under wrong skewness of the OLS residuals.

Consider the following simple data generating process:
\begin{equation}
  y=\beta_0+\sum\limits^3_{j=1}\beta_jx_{j}+\sum_{j=1}^d\delta_j z_{j}+\epsilon,
\end{equation}
with $\epsilon=v-u$ the classic composite SFA error. Here we have $p=3$ relevant inputs of production, $x_{1}$, $x_{2}$ and $x_{3}$, and $d$ environmental variables $z_{j}, j=1,\ldots,d,$ which are intended to capture variation in $y$ that can be attributed to differences in technological frontiers. 

We start with the example that all of $z_j$'s are irrelevant in the sense that $\delta_j=0$, $\forall j$.  We draw random samples from this data generating process for $n\in\{100, 200, 400, 800, 1600\}$ and we set $d=cn$ where $c\in\{0, 0.01, 0.1, 0.2, 0.3, 0.5, 0.9\}$. Thus, $c$ controls how many irrelevant variables we include in the data generating process. For $c=0$ the model is correctly and precisely specified (in the sense that it does not include irrelevant variables) while for $c>0$ we add in irrelevant variables, the number of which depends on the sample size. Note that for $n=1,600$ and $c=0.9$ we include 1,440 irrelevant covariates. 

If we think of our covariates as already in logarithmic form, then setting $\beta_0=1$, $\beta_1=0.3$, $\beta_2=0.4$ and $\beta_3=0.38$, we have a production process characterized by increasing returns to scale. Finally, we have $v\sim N(0, \sigma^2_v)$ with $\sigma_v=0.5$ and $u\sim N_+(0,\sigma^2_u)$ with $\sigma_u=1.2$ so that there is a substantial amount of inefficiency signal relative to noise. The true skewness of $\epsilon$ is $\frac{\mu_3}{\mu_2^{3/2}}= \frac{\pi-4}{\pi}\sqrt{\frac{2}{\pi}}\sigma_u^3 \left(\sigma_v^2 + \frac{\pi-2}{\pi}\sigma_u^2\right)^{-3/2} \approx -0.554$.

What is the effect of all these other variables from the standpoint of learning about inefficiency? As alluded to in Section 
\ref{sec:cols}, one of the main ways of assessing inefficiency is through skewness of the composite error. This is so because $v$ is commonly assumed to be symmetric and $u$ is one-sided, and hence this leads to the composite error term being asymmetric; the more asymmetric, the higher is expected inefficiency. However, in our setup above, as we add in more irrelevant variables, and apply OLS, we see that the average estimated skewness of the residuals goes to zero as $c\rightarrow1$. So even when COLS is feasible it is not very useful for identifying the existing inefficiency. Tables \ref{tab:skew1} and \ref{tab:skew2} illustrate this point on average, over 1,000 Monte Carlo simulations. 

\begin{table}[!h]
\caption{Average of skewness of OLS residuals over 1,000 Monte Carlo replications.  \label{tab:skew1}}
\begin{center}
\begin{tabular}{l|ccccccc}
\hline\hline
$n$&c=0&0.01&0.1&0.2&0.3&0.5&0.9\\
\hline
100&$-0.494$&$-0.488$&$-0.420$&$-0.342$&$-0.267$&$-0.143$&$-0.001$\\
200&$-0.525$&$-0.517$&$-0.445$&$-0.375$&$-0.299$&$-0.177$&$-0.011$\\
400&$-0.536$&$-0.530$&$-0.454$&$-0.380$&$-0.308$&$-0.186$&$-0.012$\\
800&$-0.547$&$-0.539$&$-0.466$&$-0.391$&$-0.319$&$-0.193$&$-0.016$\\
1,600&$-0.549$&$-0.542$&$-0.468$&$-0.391$&$-0.319$&$-0.189$&$-0.016$\\
\hline\hline
\end{tabular}\end{center}
\footnotesize{Note: The columns show the proportion of irrelevant covariates (generated as independent standard Normal) that are added to the regression based on the sample size (e.g., $0.5\times 400=200$ irrelevant variables added).}
\end{table}

Table \ref{tab:skew1} presents the average skewness for each combination of $(n,c)$. We see in the correctly specified case ($d=0$) that skewness is around -0.5, which is close to the true value,  but quickly decays to zero as $c$ increases. More troubling is that as $n$ increases, for an increasing $c$ this problem does not resolve. It is clear that if inefficiency were of primary interest, then the inclusion of irrelevant variables serves to `weaken' or `confuse' the ability of the model to correctly identify the magnitude of inefficiency that exists across firms.

To further illustrate this, Table \ref{tab:skew2} presents the proportion of samples for each combination out of the 1,000 Monte Carlo simulations that possesses the so-called wrong skewness. In these cases it is well known that both MLE and COLS estimate $\sigma^2_u$ as 0 in the Normal-Half Normal setting \cite[see][]{olson/eta:80, Waldman:82}. We see that as the proportion of irrelevant variables increases so too does the proportion of samples that generates a wrong skew. Again, for this setting, it is clear that the inclusion of irrelevant variables only serves to make it harder to correctly identify the presence of inefficiency. 

\begin{table}[!h]
\caption{Number of trials with wrong skewness of OLS residuals over 1,000 Monte Carlo replications. \label{tab:skew2}}
\begin{center}
\begin{tabular}{l|ccccccc}
\hline\hline
$n$&c=0&0.01&0.1&0.2&0.3&0.5&0.9\\
\hline
100 &$19$&$23$&$36$&$64$&$135$&$266$&$500$\\
200 &$ 1$&$ 1$&$ 7$&$10$&$ 45$&$142$&$478$\\
400 &$ 0$&$ 0$&$ 0$&$ 1$&$ 4$&$ 69$&$459$\\
800 &$ 0$&$ 0$&$ 0$&$ 0$&$ 0$&$ 8$&$416$\\
1,600&$ 0$&$ 0$&$ 0$&$ 0$&$ 0$&$ 0$&$405$\\
\hline\hline
\end{tabular}\end{center}
\footnotesize{Note: The columns show the proportion of irrelevant covariates (generated as independent standard Normal) that are added to the regression based on the sample size (e.g., $0.5 \times 400=200$ irrelevant variables added).}
\end{table}

What this simple simulation exercise illustrates is that when a practitioner includes a large number of (potentially) irrelevant covariates in a stochastic frontier model, it is more difficult to detect the true level of average inefficiency because the likelihood of wrongly skewed residuals increases and the magnitude of the skewness is on average too small. These difficulties can be mitigated by using penalized estimators which we describe next.

\subsection{LASSO}

The least absolute shrinkage and selection operator (LASSO) was initially designed as an estimation and variable selection tool for high-dimensional OLS permitting more regressors than observations \citep{tibshirani:96, knight/fu:00}. There are a few papers extending this methodology to MLE \cite[see, e.g.,][]{fan/li:01, meier/etal:08, belloni2016, jin/lee:18}. However, these extensions do not consider an estimation of skewed error distributions. 
The basic LASSO estimator adds an $\ell_1$ penalty to the sum of squared errors:
\begin{equation}\label{eq:lasso}
\widehat{\theta}^{lasso}:= \arg\min_{\theta} ||\mathbf{Y}-\mathbf{W}\theta||_{2}^{2} +\lambda ||\theta||_1, 
\end{equation}
where $||\theta||_1=\sum_{j=1}^{d}|\theta_{j}|$, and $\mathbf{W}=[\mathbf{X}, \mathbf{Z}]$ is the $n\times(p+d)$ matrix composed of $\mathbf{X}$ (the $n\times p$ matrix of production inputs) and $\mathbf{Z}$  (the $n\times d$ matrix of additional covariates) and $\mathbf{Y}$ is the $n$ vector of output measurements.

The version of the LASSO estimator we consider is defined as follows:
\begin{equation}\label{eq:lasso}
(\widehat{\beta}^{lasso}, \widehat{\delta}^{lasso}) := \arg\min_{\beta,\delta} ||\mathbf{Y}-\mathbf{X}\beta-\mathbf{Z}\delta||_{2}^{2} +\lambda ||\delta||_1, 
\end{equation}
where $||\delta||_1=\sum_{j=1}^{d}|\delta_{j}|$. Note that this is slightly different when compared to the standard LASSO from \cite{tibshirani:96} just described: this estimator forces the production inputs $\mathbf{X}$ to always remain in the model by excluding $\beta$ from the penalty part. That is, covariate selection operates only on the environmental variables $\mathbf{Z}$. This is because, in practice, a researcher usually knows which inputs are used in the production process of the industry they want to analyze. The $l_1$ penalty forces some elements of $\delta$ to be exactly zero which effectively implements variable selection and makes this estimator feasible even in cases when $d+p \ge n$.

To demonstrate that LASSO can preserve inefficiency in the stochastic frontier model, we now consider LASSO in place of OLS for the COLS estimator. That is, we run the estimator in (\ref{eq:lasso}) and evaluate the skewness of the corresponding residuals. 

Tables \ref{tab:skew1lasso} and \ref{tab:skew2lasso} present the same information as Tables \ref{tab:skew1} and \ref{tab:skew2}, but using LASSO as in (\ref{eq:lasso}) with $\lambda_{1se}$, which is the largest value of $\lambda$ selected in $10$-fold cross-validation such that the error is within one standard error of the minimum. 
Here a much different picture emerges about inefficiency. We do see that as $c$ increases there is a decline in the average estimated skewness, but this movement is not to zero, but a stable number more closely in line with the true level of skewness (Table \ref{tab:skew1lasso}). LASSO drops the irrelevant variables that obfuscate the estimation of inefficiency. Perhaps even more importantly, the proportion of the simulations with the wrong skew is virtually zero in this case (Table \ref{tab:skew2lasso}).

\begin{table}[!h]
\caption{Average of skewness of LASSO residuals over 1,000 Monte Carlo replications.  \label{tab:skew1lasso}}
\begin{center}
\begin{tabular}{l|ccccccc}
\hline\hline
$n$&c=0&0.01&0.1&0.2&0.3&0.5&0.9\\
\hline
100&$-0.503$&$-0.404$&$-0.386$&$-0.374$&$-0.367$&$-0.359$&$-0.350$\\
200&$-0.520$&$-0.452$&$-0.436$&$-0.430$&$-0.425$&$-0.420$&$-0.413$\\
400&$-0.536$&$-0.479$&$-0.470$&$-0.465$&$-0.463$&$-0.459$&$-0.455$\\
800&$-0.546$&$-0.506$&$-0.500$&$-0.498$&$-0.497$&$-0.494$&$-0.492$\\
1,600&$-0.552$&$-0.522$&$-0.519$&$-0.517$&$-0.516$&$-0.516$&$-0.514$\\
\hline\hline
\end{tabular}\end{center}
\footnotesize{Note: The columns show the proportion of irrelevant covariates (generated as independent standard Normal) that are added to the regression based on the sample size (e.g., $0.5 \times 400=200$ irrelevant variables added).}
\end{table}

For small sample sizes, even with $c=0.9$ there are only a few Monte Carlo trials that result in the wrong skew in the case of LASSO, whereas this is around 30\% when using OLS. This suggests that the idea of including variables in the hopes of explaining inefficiency is likely to work, but for the wrong reasons -- we can find spurious evidence of full efficiency (no inefficiency) as covariates unrelated to inefficiency ($u$) and to outputs ($y$) are used to proxy their variation. Indeed, incorporating a variable that explains inefficiency or output (or both) into the model should be undertaken, but if a researcher includes many variables which might spuriously correlate with inefficiency, especially in large proportions, then it should come as no surprise that such inefficiency has been `explained' away through masking it with noise. This serves to highlight the potential for machine learning methods to help in the study of inefficiency.


\begin{table}[!h]
\caption{Number of trials with wrong skewness of LASSO residuals over 1,000 Monte Carlo replications.  \label{tab:skew2lasso}}
\begin{center}
\begin{tabular}{l|ccccccc}
\hline\hline
$n$&c=0&0.01&0.1&0.2&0.3&0.5&0.9\\
\hline
100&$24$&$41$&$52$&$54$&$60$&$60$&$67$\\
200&$ 1$&$ 3$&$ 5$&$ 8$&$ 9$&$11$&$12$\\
400&$ 0$&$ 0$&$ 0$&$ 0$&$ 0$&$ 0$&$ 0$\\
800&$ 0$&$ 0$&$ 0$&$ 0$&$ 0$&$ 0$&$ 0$\\
1,600&$ 0$&$ 0$&$ 0$&$ 0$&$ 0$&$ 0$&$ 0$\\
\hline\hline
\end{tabular}\end{center}
\footnotesize{Note: The columns show the proportion of irrelevant covariates (generated as independent standard Normal) that are added to the regression based on the sample size (e.g., $0.5 \times 400=200$ irrelevant variables added).}
\end{table}



\subsection{LASSO as Selection}

While we demonstrated that LASSO preserves (as best it can) asymmetry in the composite error term in the presence of an increasing number of irrelevant variables, LASSO is more often used to select those variables that matter. In our setting this would correspond to having some (but certainly not all) of the set of $\mathbf{Z}$ variables that matter. In this case we would use LASSO to select those variables that matter (along with the variables that we have already dictated belong to the model -- our production inputs) and then estimate the model using COLS or MLE with these selected variables. This is what is known as the Post Single LASSO.

\subsubsection{Post Single LASSO}

The Post Single LASSO (PSL) approach for the stochastic frontier model is defined as a two-step procedure where we first run the LASSO as in (\ref{eq:lasso}) and then run a second-step estimator (either COLS or MLE) where we use only those $z_j$'s that have been selected by the LASSO. If we were to use COLS in the second step, the estimator can be written as: 
\begin{equation}
\left\{\widehat{\beta}^{PSL}, \widehat{\delta}^{PSL}\right\} := \arg\min_{\beta,\ \delta} ||\mathbf{Y}-\mathbf{X}\beta-\mathbf{Z}\delta||_{2}^{2}, \text{ s.t. } \delta_j = 0 \text{ for any }j \notin \text{supp}(\widehat{\delta}^{lasso}), \label{PSL}
\end{equation}
where supp of a vector is the set that identifies all non-zero elements of that vector. The residuals from Equation \eqref{PSL} are then used in the moment conditions as described in Section \ref{sec:cols}  We call this procedure PSL-COLS.

If we were instead to run maximum likelihood after use of the LASSO, we would optimize 
\begin{equation}\label{PSL_MLE}
\underset{\Theta}{\max}\sum\limits^n_{i=1}\ell(\epsilon_i;\Theta)
\end{equation}
where $\epsilon_i=y_i-\bm x_i\beta-\bm z_i \delta$ s.t. $\delta_j=0$ for any $j\notin supp (\hat\delta^{lasso})$. We will call this estimator PSL-MLE. 

For scalar $\beta$, \cite{belloni:13} demonstrate that the actual distribution of the PSL estimator of $\beta$ may be substantially biased  and that subsequent inference on $\beta$ may be invalid. 

Alternative procedures have been proposed to obtain valid inference (as we describe in the next sub-sections). The bias comes from the fact that the LASSO may not select all relevant elements of  $\mathbf{Z}$ in the first step, causing an omitted variable problem in the final step. 
However, little is known about the behavior of PSL in the setting with asymmetric errors. As an alternative to the PSL estimator, \cite{belloni:13,belloni:14} recommend the Post Double LASSO, which we now detail.

\subsubsection{Post Double LASSO}\label{sec:def:pds}

The Post Double LASSO (PDL) is arguably the most popular alternative to PSL in econometrics \cite[see, e.g.,][]{belloni:13,belloni:14, cherno/etal:18, cherno/etal:17, cherno/etal:22, belloni/etal:2013, belloni/cherno/hans/kozbur:16, belloni2016}.\footnote{Recently, \cite{wuthrich/zhu:21} argued that PDL may still fail to select all relevant variables, thus also causing an omitted variable bias in finite samples. Their simulations show that this double under-selection happens even in settings often assumed to be favorable for LASSO performance, e.g., normal and homoskedastic errors. In cases of double under-selection, their ultimate solution is to resort to OLS-based procedures (if feasible) that make inference robust to the presence of many regressors \cite[e.g.,][]{kline/etal:20}. However, their analysis does not consider distributional parameters of an asymmetric error either.} Originally developed within the framework of scalar treatment effect estimation, the idea generalizes to an arbitrary, but fixed, dimension of $\beta$. 

In a nutshell, the idea is to carry out two separate LASSO stages, one for $Y$ on $\mathbf{Z}$ (without $\mathbf{X}$) and the other for each variable in $\mathbf{X}$ on $\mathbf{Z}$. 
Therefore, the PDL is defined as a three-step procedure. The first step is for LASSO to select relevant variables from $\mathbf{Z}$ that predict $\mathbf{Y}$. The second step is for LASSO to select relevant variables from $\mathbf{Z}$ that predict, one by one, each regressor $\mathbf{X}_l, l=1, \ldots, p$. The third step is to deploy OLS of $\mathbf{Y}$ on $\mathbf{X}$ and \emph{all} the selected elements of $\mathbf{Z}$ from the prior LASSO steps. We can write the estimator as follows:
\begin{eqnarray}
1:&\qquad \widehat{\pi}_1^{lasso} := & \arg\min_{\pi_1} ||\mathbf{Y}-\mathbf{Z}\pi_1||_{2}^{2} +\lambda_1 ||\pi_{1}||_1 \\
2:&\qquad \widehat{\pi}_{1+l}^{lasso} := & \arg\min_{\pi_{1+l}} ||\mathbf{X}_l-\mathbf{Z}\pi_{1+l}||_{2}^{2} +\lambda_{1+l} ||\pi_{1+l}||_1, \quad l=1,\ldots, p\\ \label{PDL}
3: &\left\{\widehat{\beta}^{PDL}, \widehat{\delta}^{PDL}\right\} := & \arg\min_{\beta,\ \delta} ||\mathbf{Y}-\mathbf{X}\beta-\mathbf{Z}\delta||_{2}^{2}, \label{eq:PDL-COLS}\\\nonumber
&&\text{ s.t. } \delta_j = 0 \text{ for any }j \notin \hat{I}=\hat{I}_1 \cup \hat{I}_2\cup\ldots \hat{I}_{1+p},
\end{eqnarray}
where $\hat{I}_1 =\text{supp}(\widehat{\pi}_1^{lasso})$ and $\hat{I}_{1+l} =\text{supp}(\widehat{\pi}_{1+l}^{lasso}), l=1, \ldots, p$.

In the context of the stochastic frontier model, the PDL for maximum likelihood estimation would simply replace OLS estimation in Step 3 above (Equation \eqref{eq:PDL-COLS}) with maximum likelihood as in Equation \eqref{PSL_MLE}. We call the two resulting estimators PDL-COLS and PDL-MLE, respectively. 

An important component of debiasing that happens in PDL is achieved through cross-fitting \cite[see, e.g.,][]{newey/robins:18}. This is when we  split the data into two subsets and estimate $\pi_1, \ldots, \pi_{p+1}$ using one subset, and $\beta, \delta$ using the other. This ensures that each observation is never used for both the LASSO steps and the OLS/MLE step simultaneously, reducing correlation between the estimation errors in the two sets of estimates. We note that cross-fitting is done in addition to any sample splitting (cross-validation) used to determine the optimal value of hyperparameters $\lambda_1, \ldots, \lambda_{p+1}$.







\subsection{PSL and PDL for the Stochastic Frontier Model}

A more realistic setting to study the impact of inclusion of potentially many irrelevant covariates is when some but not all variables from $\mathbf{Z}$ are irrelevant and the sample size is not large enough for LASSO to reliably select the relevant covariates. This is the setting where PSL-OLS is known to produce biased estimates of the slope coefficients \cite[see, e.g.,][]{belloni:14}. How does this bias play out for COLS and MLE?

We follow Design 1 of \citet[][Section 6]{belloni:13}. We generate 200 $z_j$'s which are Normal covariates with covariance $0.5^{|k-l|}$ where $k$ and $l$ are the indexes of any two of them, $k\ne l \in\{1, \ldots, 200\}$. For simplicity, we consider one input $x$. The DGP is as follows: 
\begin{eqnarray}\label{pdl:y}
y&=&\beta_0+\beta x+\sum_{j=1}^{200} c_y\delta_j z_{j}+v-u\\\label{pdl:x}
x&=&\sum_{j=1}^{200}c_x\delta_j z_{j}+\eta,
\end{eqnarray}
where $v\sim N(0, \sigma^2_v)$ with $\sigma^2_v=0.5$, $u\sim N_+(0,\sigma^2_u)$ with $\sigma^2_u=1.2$ and $\eta\sim N(0,1)$. The true values of $\delta_j$ are $(1/j)^2$ and the constants $c_x=0.8$ and $c_y=0.6$ are picked to match the values used by \cite{belloni:14}. The sample size is 100. 

\begin{figure}[!h]
\begin{center}
\caption{Sampling distribution of Post Single LASSO COLS (PSL-COLS) and Post Double LASSO COLS (PDL-COLS) estimators.} \label{fig:PSL_PDL_COLS}
\includegraphics[width=\textwidth
]{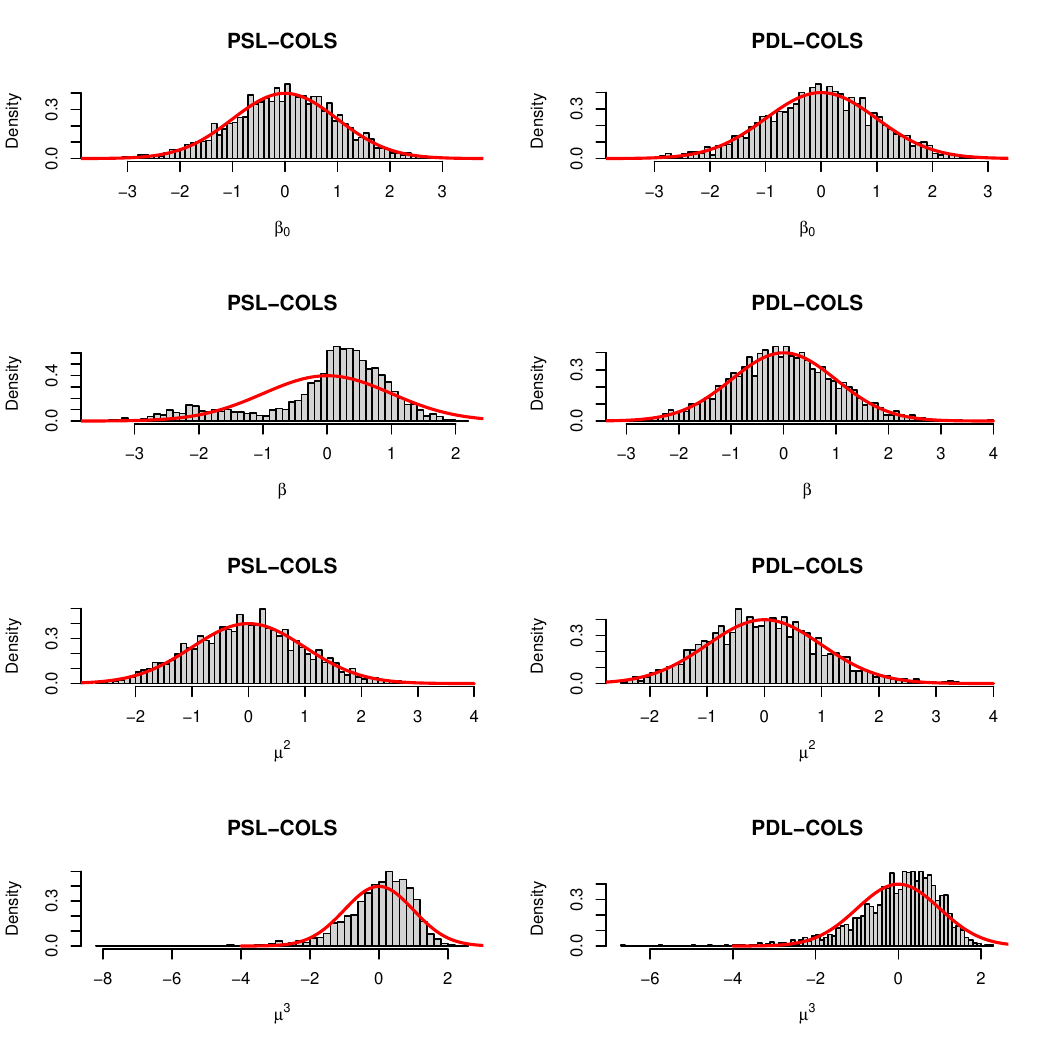}
\end{center}
\end{figure}

\begin{figure}[!h]
\begin{center}
\caption{Sampling distribution of Post Single LASSO MLE (PSL-MLE) and Post Double LASSO MLE (PDL-MLE) estimators.} \label{fig:PSL_PDL_MLE}
\includegraphics[width=\textwidth
]{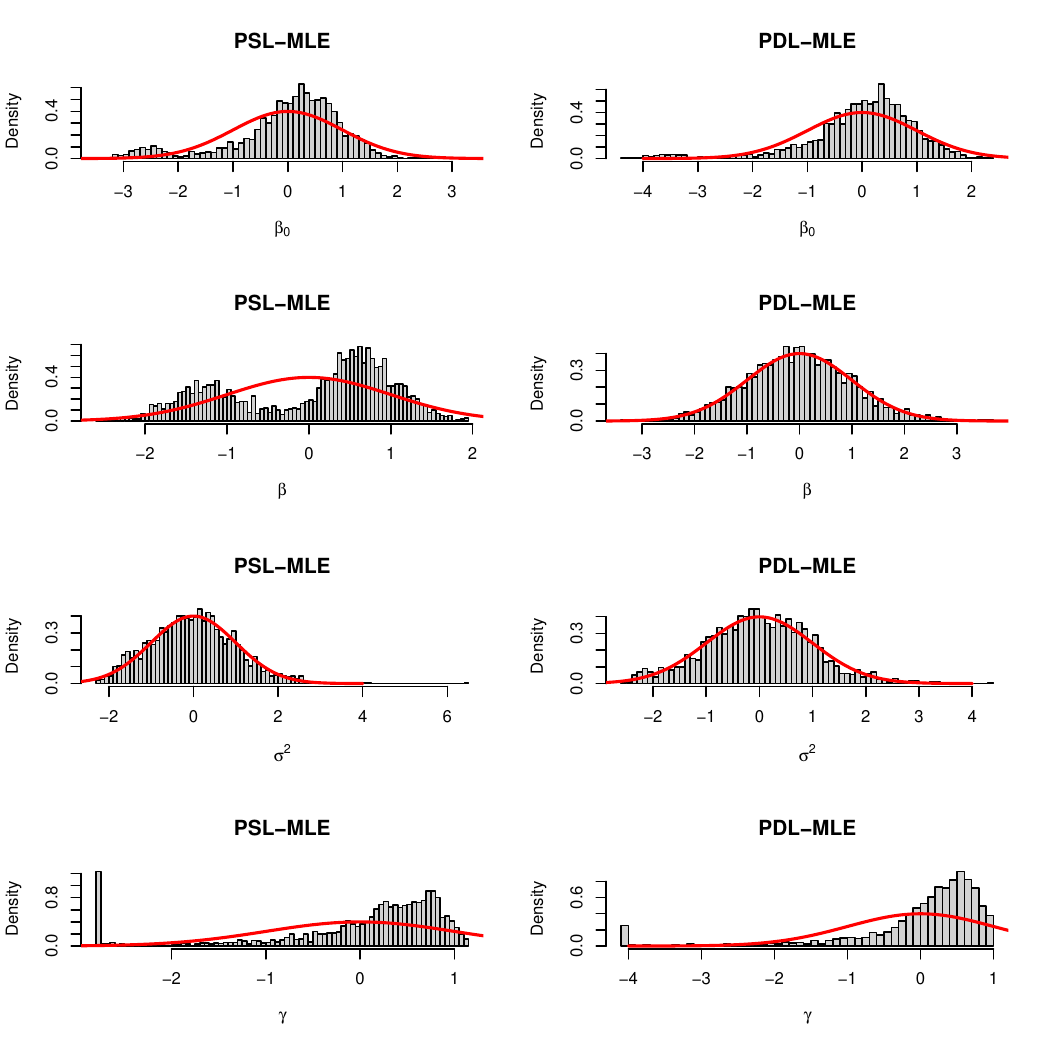}
\end{center}
\end{figure}

It is clear from this DGP that the magnitude of the $\delta_j$'s decreases at an exponential rate so that only the first few elements of $\mathbf{Z}$ are relevant. Given the sample size, COLS and MLE are infeasible and we have to run PSL-COLS/PDL-COLS or PSL-MLE/PDL-MLE. Arguably this setup is the least favorable setting for LASSO because $\delta_j$'s can be small but never equal zero. We report 
the scaled and centered sampling distributions of both estimators in Figures \ref{fig:PSL_PDL_COLS} (for COLS) and \ref{fig:PSL_PDL_MLE} (for MLE) as histograms over 2,000 replications against the standard Normal density (red curve). 

It is evident from Figure \ref{fig:PSL_PDL_COLS} that the PDL-COLS eliminates the bias in $\hat{\beta}$ present in PSL-COLS and that the underselection bias of LASSO does not carry over to the estimators of $\beta_0$, $\mu_2$ or $\mu_3$. However, note that PSL-COLS suffers more than PDL-COLS from the wrong skew problem: the number of times $\hat{\mu}_3>0$ was 48 for PSL-COLS and only 14 for PDL-COLS (out of 2,000 simulations).

Similar to Figure \ref{fig:PSL_PDL_COLS}, we report in Figure \ref{fig:PSL_PDL_MLE} the scaled and centered sampling distributions of PSL-MLE and PDL-MLE for estimators of $\beta_0$, $\beta$, $\sigma^2$ and $\gamma$ over 2,000 replications. We observe a similar underselection bias in the estimation of $\beta$ as for PSL-COLS which is remedied by double selection PDL-COLS. Also, we observe the spike in PSL-MLE of $\gamma$ corresponding to values near zero. The spike is lower for PDL-MLE. This spike in $\gamma$ has direct consequences for the estimation of inefficiency for a given sample.

\begin{figure}[!h]
\begin{center}
\caption{Sampling distribution of mean efficiency.} \label{fig:u}
\includegraphics[width=0.8\textwidth, height=6cm
]{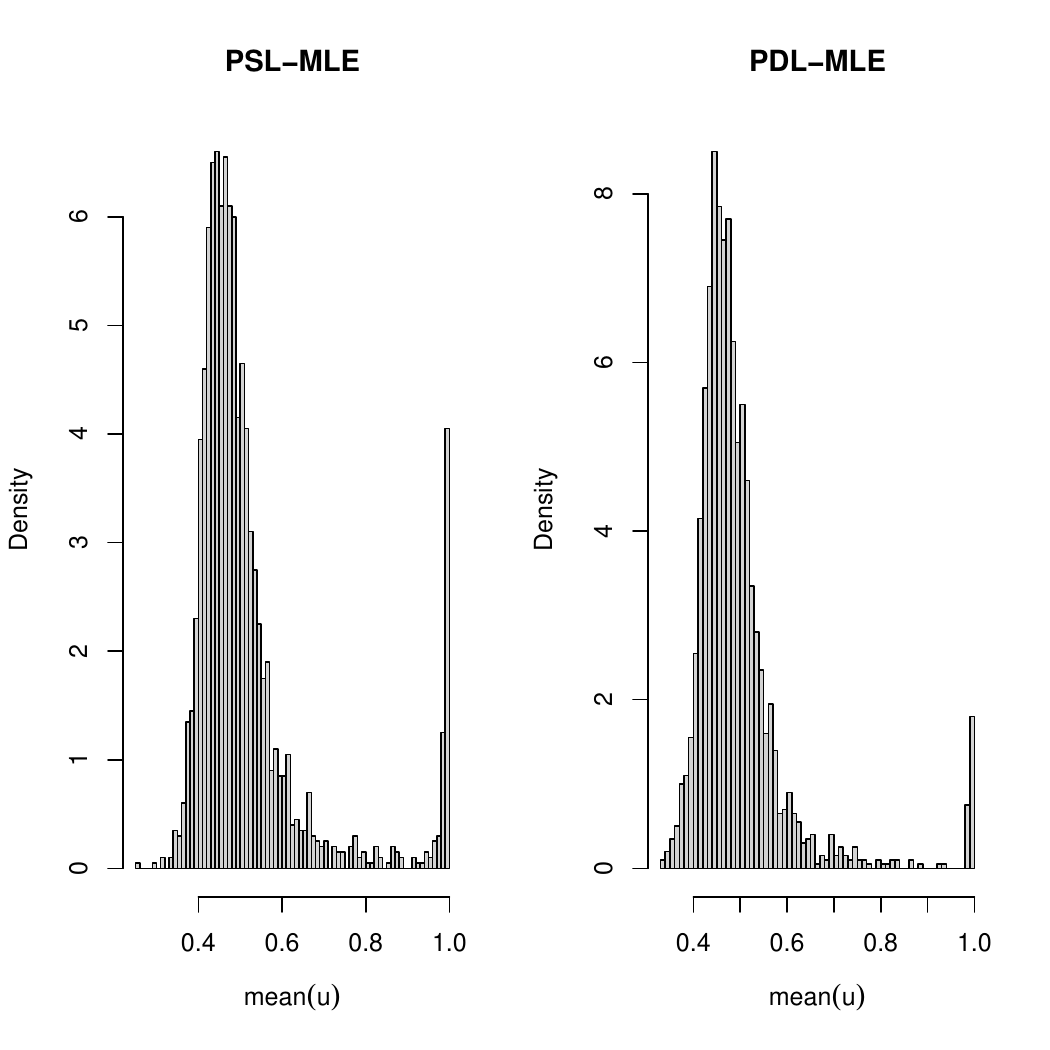}
\end{center}
\end{figure}

This is presented in Figure \ref{fig:u} which shows the sampling distribution of the mean efficiency across the 2,000 simulations for the PSL and PDL MLE estimator. One can see clearly that the frequency of samples that find no inefficiency is substantially higher in PSL than in PDL.

\section{Neyman Orthogonality for the Stochastic Frontier Model}\label{sec:NO}

In this section we briefly outline the theoretical justifications for the PDL applied to the stochastic frontier model. Basically, the reason PDL works is that (under certain assumptions) it is able to eliminate the omitted variable bias of PSL.

\subsection{COLS}

Consider the COLS estimator of the model in Eq.~(\ref{eq:sfa}) and assume for simplicity that $x_i$ and $z_i$ are scalars. The moment conditions solved by OLS for $\beta$ and $\delta$ can be written as follows: 
\[\begin{array}{cc}
\text{[A] }& \mathbb{E} x_i\epsilon_i=0 \\   
\text{[B] }& \mathbb{E} z_i\epsilon_i= 0,   
\end{array} \]
where $\epsilon_i=y_i - \beta_0-x_i \beta - z_i\delta$. LASSO produces a regularization bias so that $\mathbb{E}\widehat{\delta}^{lasso} \ne \delta_0$, where $\delta_0$ is the true value of $\delta$. So using $\widehat{\delta}^{lasso}$ in the second step of PSL-COLS is equivalent to using the following moment condition instead of [A]:
\[\begin{array}{cc}
[\text{A}^\prime] & \mathbb{E} x_i\left(y_i - \beta_0-x_i \beta - z_i\widehat{\delta}^{lasso} \right)=0 
\end{array} \]
Moment condition [\text{A}$^\prime$] is invalid. Indeed, this moment function can be written as $\mathbb{E}x_i\left(y_i - \beta_0-x_i \beta - z_i\delta_0\right) + \mathbb{E}x_iz_i(\delta_0-\widehat{\delta}^{lasso})$ which is not mean-zero. 
This gives rise to an omitted variable bias unless $\mathbb{E}x_i z_i=0$. An easy way to eliminate the bias is to use the part of $x_i$ orthogonal to $z_i$ instead of $x_i$. This will turn the second summand, $\mathbb{E}x_iz_i(\delta_0-\widehat{\delta}^{lasso})$, into zero. 

As a result of using the linear error in the projection of $x_i$ on $z_i$, the new  moment condition becomes
\[\begin{array}{cc}
[\text{A}^{\prime\prime}]& \mathbb{E} x^{\perp}_i\left(y_i - \beta_0-x_i \beta - z_i\widehat{\delta}^{lasso} \right)=0, 
\end{array} \]
where $x^\perp_i=(x_i-\pi_x z_i)$ and $\pi_x = \mathbb{E} (x_i z_i)[\mathbb{E} (z_i^2)]^{-1}$. This moment condition is valid regardless of how far $\widehat{\delta}^{lasso} $ is from $\delta_0$. To see this, note that
\[\mathbb{E} x^\perp_i z_i\widehat{\delta}^{lasso} =\mathbb{E} \left[x_iz_i-\mathbb{E} (x_i z_i)\mathbb{E} (z_i^2)^{-1} z_i^2\right]\widehat{\delta}^{lasso}=\left[\mathbb{E} (x_iz_i)-\mathbb{E} (x_i z_i)\right]\widehat{\delta}^{lasso}= 0.\]

Moreover, the distribution of the estimator of $\beta$ is insensitive to the behavior of $\widehat{\delta}^{lasso}$ because the expected derivative of this moment function with respect to $\delta$ is zero. This can be seen by writing
\[\begin{array}{cc}
\text{ }& \mathbb{E}\frac{\partial}{\partial\delta} x^\perp_i\left(y_i - \beta_0-x_i \beta - z_i\widehat{\delta}^{lasso} \right)=\mathbb{E}(-z_i)(x_i-\pi_x z_i)=0,
\end{array} \]
which is the moment condition that would be used to estimate $\pi_x$. 

However, if we use regularization in estimating $\pi_x$, then $\widehat\pi_x^{lasso}$ is biased, which invalidates moment condition [A$^{\prime\prime}$]. We can apply a similar trick to restore the validity of the moment condition by considering the part of $\epsilon_i$ orthogonal to $z_i$. The moment condition becomes
\[\begin{array}{cc}
\text{[A*] }& \mathbb{E} x^\perp_i\left[y^\perp_i - \beta_0-x^\perp_i\beta -z_i\delta \right]=0,
\end{array} \]
where $y^\perp_i=y_i-\pi_yz_i$ and the estimation of $\beta$ is insensitive to the behavior of $\widehat{\delta}^{lasso}$, $\widehat{\pi}_x^{lasso}$, and $\widehat{\pi}_y^{lasso}$. Due to this property, moment condition [A*] is called Neyman orthogonal. This is also what \cite{PROKHOROV/schmidt:09} call \emph{moment and parameter (M/P) redundancy}, which is when neither the knowledge of the additional moment conditions nor the knowledge of the parameters they identify ($\delta$, $\pi_x$, $\pi_y$) help improve efficiency of estimation of $\beta$ \cite[see also][Section 2.3]{hirukawa/etal:23}. The PDL estimator $\widehat\beta^{PDL}$ satisfies it.

\subsection{Neyman Orthogonality for MLE}
We now turn to MLE. Assume for ease of exposition that $\sigma=\lambda=1$ and consider the moment conditions that are implied by the MLE of $\beta$ and $\delta$. These moment conditions correspond to the relevant first-order conditions of MLE derived by \citet[][Eqs. (11)-(13)]{ALS:77}: 
\[\begin{array}{cc}
\text{[A] }& \mathbb{E} x_i\left(\epsilon_i + \frac{\phi_i}{1-\Phi_i}\right)=0 \\   
\text{[B] }& \mathbb{E} z_i\left(\epsilon_i + \frac{\phi_i}{1-\Phi_i}\right)= 0   
\end{array} \]
where $\phi_i:=\phi\left(\epsilon_i\right)$ and $\Phi_i:=\Phi\left(\epsilon_i\right)$ are the standard normal pdf and cdf, respectively, evaluated at $\epsilon_i=y_i - \beta_0-x_i \beta - z_i\delta$. The full set of first-order conditions include one more moment for each of $\lambda$ and $\sigma$ but we leave this aside for the moment. What is the Neyman orthogonal moment condition here?

Following the reasoning of the previous subsection, a natural approach is to use in moment conditions [A] and [B] the parts of $y_i$ and $x_i$ orthogonal to $z_i$. Let $\epsilon^*_i := y^\perp_i - \beta_0-x^\perp_i\beta -z_i\delta$ and let $\phi^*_i:=\phi\left(\epsilon^*_i\right)$ and $\Phi^*_i:=\Phi\left(\epsilon^*_i\right)$. Then, we can define the inverse Mills ratio as $r^*_i:=\phi^*_i/(1-\Phi^*_i)$. We note that conditional homoskedasticity of $\epsilon_i|x_i, z_i$ implies conditional homoskedasticity of $\epsilon^*_i|x_i, z_i$ and of $r^*_i|x_i, z_i$. That is, $\mathbb{E}{\epsilon^*_i}^2|x_i, z_i$, $ \mathbb{E}{r^*_i}^2|x_i, z_i$, and $\mathbb{E}\epsilon^*_ir^*_i|x_i, z_i$ are scalars independent of $x_i, z_i$. 
Consider the moment conditions
\[\begin{array}{cc}
\text{[A*] }& \mathbb{E} x^\perp_i\left(\epsilon^*_i + r^*_i\right)=0 \\   
\text{[B*] }& \mathbb{E} z_i\left(\epsilon^*_i + r^*_i\right)= 0,   
\end{array} \]
such that $\mathbb{E} z_i x^\perp_i=0$. These moment conditions correspond to running MLE where the dependent variable is the part of $y_i$ orthogonal to $z_i$ and the explanatory variables are $z_i$ and the part of $x_{i}$ orthogonal to $z_i$, i.e.~$x^\perp_i$. 

To see that [A*] satisfies Neyman orthogonality we compute expected derivatives with respect to all the nuisance parameters: 
\[\begin{array}{cl}
\delta: & \mathbb{E} x^\perp_i\left(\frac{\partial}{\partial\delta}\epsilon^*_i + \frac{\partial}{\partial\delta} r^*_i \right)=\mathbb{E}x^\perp_i(-z_i + z_i r^*_i(\epsilon_i^*-r^*_i))=0,\\
\pi_x:& \mathbb{E} (-z_i)(\epsilon^*_i + r^*_i )+\mathbb{E}(x_i-\pi_x z_i)\left(\frac{\partial}{\partial\pi_x}\epsilon^*_i + \frac{\partial}{\partial\pi_x} r^*_i \right)=0+ \mathbb{E}x^\perp_i(-z_i + z_i r^*_i(\epsilon_i^*-r^*_i))\beta=0,\\
\pi_y:& \mathbb{E} x^\perp_i\left(\frac{\partial}{\partial\pi_y}\epsilon^*_i + \frac{\partial}{\partial\pi_y} r^*_i \right)=\mathbb{E}x^\perp_i(-z_i + z_i r^*_i(\epsilon_i^*-r^*_i))=0,
\end{array} \] 
where $\frac{\partial}{\partial\delta} r^*_i=\frac{1}{\beta}\frac{\partial}{\partial\pi_x} r^*_i=\frac{\partial}{\partial\pi_y} r^*_i=z_i r^*_i(\epsilon_i^*-r^*_i)$ by properties of the inverse Mills ratio, $\mathbb{E} (-z_i)(\epsilon^*_i + r^*_i )=0$ by [B*]. To see that $\mathbb{E}x^\perp_i z_i r^*_i(\epsilon_i^*-r^*_i)=0$, apply the law of iterated expectations at the true value of $\pi_x$ as follows 
\[\mathbb{E}x^\perp_i z_i r^*_i(\epsilon_i^*-r^*_i)=\mathbb{E}_{x,z}z_i x^\perp_i \left(\mathbb{E}r^*_i\epsilon_i^*|x_i, z_i-\mathbb{E}{r^*_i}^2|x_i, z_i\right).\] 
Now the two conditional terms are fixed scalars and can be moved outside the conditional expectation operator. It remains to notice that $\mathbb{E}z_ix^\perp_i = \mathbb{E}z_i(x_i-\pi_x z_i)=0$, which is the moment identifying $\pi_x$.

Therefore, the estimator based on [A*] is not affected by regularization biases or other aspects of the sampling distributions of $\widehat{\delta}^{lasso}$, $\widehat{\pi}_x^{lasso}$, and $\widehat{\pi}_y^{lasso}$. Finally, it is worth noting that the above results are also robust to deviations from the distributional assumptions of MLE, as long as the moment conditions [A] and [B] hold, which can be true more generally, not only for the Normal/Half-Normal error term.

\section{Empirical Application}

To illustrate the method, we use data from \cite{APS:2016}, which contains information on 137 farms from Northern Spain over 12 years (1999-2010).\footnote{A balanced panel for this data was also recently studied by \cite{AMSLER_SCHMIDT:2019} for different aims.} This data set is an extension of the data from \cite{ALVAREZ_ARIAS:2004} on 196 dairy farms that were participating in a voluntary extension program, where farms were visited monthly by a technician who took detailed records about technical and accounting state of the farm operations over 6 years (1993-1998). A longitudinal data set like this presents a close to ideal setting for a researcher to illustrate new methods, hence our choice, though a similar application can be carried out for other data, whether for micro or macro levels.
 
More specifically, the data set contains information on milk production (in liters), which is the single output variable dubbed hereafter as \textit{Milk}, and five inputs: \textit{Labor} (number of man-equivalent units), \textit{Cows} (number of milking cows), \textit{Feedstuffs} (total amount of feedstuffs fed to the dairy cows in kgs.), \textit{Land} (hectares of land devoted to pasture and crops), \textit{Roughage} (expenses incurred in producing roughage on the farm (in Euros), such as expenses for fertilizer, seeds and sprays, hired machinery and labor, the depreciation of the machinery). We refer to \cite{ALVAREZ_ARIAS:2004} for more details and discussion about and the importance of these variables for milk production. The extension to their data which was used in \cite{APS:2016} also contained information on landownership in terms of the percentage of land owned by the farmer (\textit{Landown}), standardized milk quality indicators in terms of bacteriological content (\textit{Germ}), fat content (\textit{Fat}), protein content (\textit{Prot}) and somatic cell count (\textit{SCC}), price of milk (\textit{PMilk}), price of feed (\textit{PFeed}), and membership in agricultural cooperative (seven dummy variables), in addition to eleven time dummies. 

Table \ref{tab:SDF_sumstat} contains the summary statistics. Nothing out of the ordinary appears from the summary statistics, though it should be highlighted that these are all small farms. 

\begin{table}[!h] \centering 
 \caption{Summary statistics for Spanish dairy farm data.} 
 \label{tab:SDF_sumstat} 
\begin{tabular}{@{\extracolsep{5pt}}lccccc} 
\hline\hline
Statistic & \multicolumn{1}{c}{Mean} & \multicolumn{1}{c}{St. Dev.} & \multicolumn{1}{c}{Min} & \multicolumn{1}{c}{Median} & \multicolumn{1}{c}{Max} \\ 
\hline
Milk & 331,845 & 207,558 & 19,037 & 275,106 & 1,322,276 \\ 
Feedstuffs & 154,243.300 & 102,811.900 & 11,406 & 126,350 & 707,627 \\ 
Cows & 42.319 & 21.198 & 4 & 38 & 134 \\ 
Land & 19.351 & 8.861 & 2 & 18 & 57 \\ 
Labor & 1.793 & 0.794 & 0.150 & 1.980 & 5.740 \\ 
Roughage & 21,105.230 & 17,255.280 & 24 & 16,272 & 132,750 \\ 
PMilk & 0.374 & 0.040 & 0.210 & 0.380 & 0.540 \\ 
PFeed & 0.271 & 0.039 & 0.130 & 0.270 & 1.060 \\ 
Landown & 0.586 & 0.258 & 0 & 0.609 & 1\\ 
Germs & 22,792.900 & 15,118.010 & 0 & 18,417 & 105,533 \\ 
Protein & 3.143 & 0.129 & 2.250 & 3.140 & 4.050 \\ 
Fat & 3.767 & 0.208 & 2.500 & 3.770 & 4.740 \\ 
SCC& 181,949.900 & 118,625 & 30 & 189,917 & 797,667 \\ 
AvgCost & 0.225 & 0.046 & 0.078 & 0.222 & 0.397 \\ 
\hline\hline
\end{tabular} 
\end{table} 

We refer to \cite{APS:2016} for more details and discussion about these additional variables and how they can influence the frontier or inefficiency for this sector. Additionally, 

We also recognize that any of these variables may enter the model in a non-linear form, possibly also with interactions with any other terms. On the one hand, not accounting for such non-linear influence may jeopardize the accuracy of the frontier estimation. On the other hand, accounting for all possible non-linear influences may simply be infeasible. What is practical, and similar to the argument in the application of \cite{belloni:14}, is to consider at least the first and second order terms (including interactions) and let the LASSO help by shrinking and selecting the most influential of these terms. All told we have 8 continuous external covariates, 12 time dummies, 8 COOP dummies, 5 zonal dummies, and 18 county dummies. This is a total of 51 first order terms.\footnote{Both Germs and SCC are standardized prior to any analysis.} 

\subsection{Cobb-Douglas with All Inputs Mandatory}

In this setting we treat the five main inputs as mandatory variables in the production process and assume a Cobb-Douglas functional form. Estimates comparing across the different approaches appear in Table \ref{tab:SDF_prod_CD}. Average efficiency and returns to scale (RTS) are also presented. For our production frontier model (estimated with either COLS or MLE), these specifications include our five covariates (in log deviations from their respective means). PSL-COLS and PSL-MLE are our post single LASSO estimators. Both of these estimates run a single LASSO with the 5 main covariates (inputs) included without selection and the 51 covariates where the selection via LASSO-penalty is done.\footnote{We use the penalty selection as advocated in \cite{belloni:14}.} After this LASSO step is undertaken, then either COLS or MLE is conducted using the five inputs and the selected subset of the 51 covariates. PDL-COLS and PDL-MLE applies the same logic but runs a LASSO for Milk production on the 51 covariates (without the five inputs), as well  as a LASSO for each of the five inputs on the 51 covariates. Then, we take the union over this set of selected covariates from the six LASSOs and run MLE or COLS on them along with the five inputs.

\begin{table}[!h] \centering 
 \caption{Estimates of Cobb-Douglas stochastic production frontier. Standard errors appear beneath each estimate in parentheses.}
 \label{tab:SDF_prod_CD} 
\begin{tabular}{l|cccccccc}
\hline\hline
&\multicolumn{2}{c}{No $Z$}&\multicolumn{2}{c}{All $Z$}&\multicolumn{2}{c}{PSL}&\multicolumn{2}{c}{PDL}\\
&\multicolumn{1}{c}{COLS}&\multicolumn{1}{c}{MLE}&\multicolumn{1}{c}{COLS}&\multicolumn{1}{c}{MLE}&\multicolumn{1}{c}{COLS}&\multicolumn{1}{c}{MLE}&\multicolumn{1}{c}{COLS}&\multicolumn{1}{c}{MLE}\\
\hline
Feedstuffs&$ 0.386$&$ 0.360$&$ 0.464$&$ 0.464$&$ 0.439$&$ 0.439$&$ 0.398$&$ 0.387$\\
&$ (0.012)$&$ (0.013)$&$ (0.011)$&$ (0.011)$&$ (0.011)$&$ (0.013)$&$ (0.012)$&$ (0.013)$\\
Cows&$ 0.595$&$ 0.642$&$ 0.464$&$ 0.464$&$ 0.546$&$ 0.546$&$ 0.555$&$ 0.571$\\
&$ (0.020)$&$ (0.022)$&$ (0.017)$&$ (0.017)$&$ (0.017)$&$ (0.021)$&$ (0.020)$&$ (0.021)$\\
Land&$-0.010$&$-0.012$&$-0.002$&$-0.002$&$ 0.007$&$ 0.007$&$ 0.003$&$ 0.004$\\
&$ (0.009)$&$ (0.009)$&$ (0.008)$&$ (0.008)$&$ (0.008)$&$ (0.010)$&$ (0.010)$&$ (0.010)$\\
Labor&$ 0.035$&$ 0.032$&$ 0.013$&$ 0.013$&$-0.015$&$-0.015$&$ 0.017$&$ 0.014$\\
&$ (0.012)$&$ (0.012)$&$ (0.010)$&$ (0.009)$&$ (0.009)$&$ (0.011)$&$ (0.011)$&$ (0.011)$\\
Roughage&$ 0.067$&$ 0.060$&$ 0.074$&$ 0.074$&$ 0.082$&$ 0.082$&$ 0.062$&$ 0.060$\\
&$ (0.005)$&$ (0.005)$&$ (0.004)$&$ (0.004)$&$ (0.004)$&$ (0.005)$&$ (0.005)$&$ (0.005)$\\
\hline
RTS&$ 1.074$&$ 1.082$&$ 1.014$&$ 1.014$&$ 1.059$&$ 1.059$&$ 1.035$&$ 1.037$\\
Mean Eff&$ 0.930$&$ 0.892$&$ 1$&$ 0.999$&$ 1$&$ 0.999$&$ 0.954$&$ 0.924$\\
Num $Z$&--&--&$51$&$51$&$ 8$&$ 8$&$26$&$26$\\
\hline\hline
\end{tabular}
\end{table} 

There are a few immediate features to notice. First of all, note that the `no $Z$' COLS and MLE yielded somewhat similar results: RTS of about 1.07-1.08, and efficiency of about 89-93\%. Also note that MLE excluding the $Z$ variables has the lowest average efficiency (0.892) but also has a bizarre negative sign on the input elasticity of land (this also happens with COLS with no $Z$ variables included).  Moreover, the estimated coefficients and their standard errors are also fairly similar between COLS and MLE, yet not the same (as one might expect due to asymptotic equivalence). For example, the coefficients for the most important inputs (Cows and Feedstuffs) are different by about two standard errors for COLS relative to MLE. This might be an indication of oversimplification of the model, due to omission of some $Z$ variables. On the other hand,     when we use all of the potential first order terms in our $Z$ matrix, then lower RTS estimates are obtained (1.01 for both COLS and MLE) and the highest average efficiency score (100\% for COLS and 99.9\% for MLE). For the COLS model including all of the $Z$ variables, the efficiency scores are all 100\% as the corresponding residuals had positive skewness which precluded identification of positive $\sigma_u$ in a Normal-Half Normal specification of the composite error. 

Furthermore, note that the estimated coefficients and their standard errors are almost identical when comparing COLS and MLE for this ``All $Z$'' case. However, they are substantially different from those without any $Z$'s.  For example, the coefficients for the most important inputs Feedstuffs and Cows are slightly under 0.4 and about 0.6, respectively, for COLS and MLE without $Z$'s, but when we include all $Z$'s both estimated coefficients are 0.464 (for both COLS and MLE). Furthermore, the coefficient for Labor becomes not significant. This might be an indication of overcomplication of the model with too many $Z$. This naturally motivates a need to  try our optimal variable selection approaches -- single and double LASSO, which we consider next. 

Interestingly, the PSL-MLE yields similar results as MLE including all of the $Z$ variables: 99\% efficiency and slightly higher RTS of 1.059, which are fairly different from the COLS and MLE excluding all of the $Z$ variables. The coefficient estimates for these models are almost identical when we compare PSL-COLS with PSL-MLE, but some are substantially different when we compare with those with or without $Z$. For example, the estimated coefficients for Cows is now about 0.546 for both PSL-COLS and PSL-MLE (with standard errors around 0.02), while they were ranged between 0.6-0.64 without $Z$'s and were 0.464 with all $Z$'s, i.e., outside two standard errors. Also, it is worth noting that PSL-MLE shows a negative sign for labor, which is also bizarre, indicating something is likely to be wrong with this model.

An interesting observation to note is that, out of the potential 51, the PSL included only 8 of $Z$'s, while PDL included 26 $Z$'s. For our PDL results we see that all of the input coefficients have positive signs (as anticipated) with scale economies of roughly 4\% (i.e., RTS$\approx$1.04) and an average farm efficiency score of 0.92-0.95. That is, the PDL-MLE and PDL-COLS yielded efficiency similar to that estimated from COLS excluding all of the $Z$ variables, but smaller estimate of RTS and, importantly, no negative coefficients. 

When we compare estimated coefficients of PDL-COLS with PDL-MLE we see that they are not identical (as was observed for the PSL case), yet very similar (as one would expect).  We also see that some of them are substantially different from those estimated without $Z$ or with all $Z$ or the PSL approaches.  For example, the estimated coefficients for Feedstuffs (one of the most important inputs) is now about 0.4 for both PDL-COLS and PDL-MLE (with standard errors around 0.01), while they were 0.46 with all $Z$'s and 0.44 with PSL, i.e., outside the two standard errors. Meanwhile, this estimate is closest to that of COLS without $Z$'s (yet substantially different from MLE without $Z$'s). On the other hand, the estimate of the coefficient for Cows in PDL is similar to that in PSL, yet substantially different from the estimates without $Z$'s (especially with MLE) and with all $Z$'s. All these differences provide a vivid illustration of the sensitivity of the results (not only efficiency and RTS but also some key coefficients of frontier itself) with respect to the choice of $Z$-variables. Of course, as this is real data, we do not know what the ``true'' coefficients are, yet we would tend to have higher confidence in the PDL results given the theoretical reasoning and the simulation evidence presented above.  

\subsection{Cobb-Douglas with Only Some Inputs Mandatory}

Next, we recognize that in some situations not all of the ``inputs'' may be obvious for inclusion, or their impact may be dubious. We consider this case for our baseline setting with land, labor and roughage. Thus, we repeat the analysis in Table \ref{tab:SDF_prod_CD}, but instead of forcing land, labor and roughage to appear, we instead move them to our set of potential $Z$ variables (increasing the number from 51 to 54). These estimates appear in Table \ref{tab:SDF_prod_CD_bigZLessX}.

\begin{table}[!h]
\caption{Estimates of Cobb-Douglas stochastic production frontier but only allowing Cows and Feedstuffs to be included by default. Standard errors appear beneath each estimate in parentheses.}
 \label{tab:SDF_prod_CD_bigZLessX}
\begin{center}
\begin{tabular}{l|cccccccc}
\hline\hline
&\multicolumn{2}{c}{No $Z$}&\multicolumn{2}{c}{All $Z$}&\multicolumn{2}{c}{PSL}&\multicolumn{2}{c}{PDL}\\
&\multicolumn{1}{c}{COLS}&\multicolumn{1}{c}{MLE}&\multicolumn{1}{c}{COLS}&\multicolumn{1}{c}{MLE}&\multicolumn{1}{c}{COLS}&\multicolumn{1}{c}{MLE}&\multicolumn{1}{c}{COLS}&\multicolumn{1}{c}{MLE}\\
\hline
Feedstuffs&$0.436$&$0.384$&$ 0.464$&$ 0.464$&$0.438$&$0.438$&$ 0.391$&$ 0.374$\\
&$(0.012)$&$(0.013)$&$ (0.011)$&$ (0.011)$&$(0.011)$&$(0.013)$&$ (0.012)$&$ (0.013)$\\
Cows&$0.677$&$0.738$&$ 0.464$&$ 0.464$&$0.541$&$0.541$&$ 0.561$&$ 0.587$\\
&$(0.017)$&$(0.018)$&$ (0.017)$&$ (0.017)$&$(0.015)$&$(0.022)$&$ (0.020)$&$ (0.022)$\\
Land&--&--&$-0.002$&$-0.002$&--&--&$ 0.012$&$ 0.012$\\
&--&--&$ (0.008)$&$ (0.008)$&--&--&$ (0.009)$&$ (0.009)$\\
Labor&--&--&$ 0.013$&$ 0.013$&--&--&$ 0.019$&$ 0.017$\\
&--&--&$ (0.010)$&$ (0.009)$&--&--&$ (0.011)$&$ (0.011)$\\
Roughage&--&--&$ 0.074$&$ 0.074$&$0.082$&$0.082$&$ 0.065$&$ 0.061$\\
&--&--&$ (0.004)$&$ (0.004)$&$(0.004)$&$(0.009)$&$ (0.005)$&$ (0.005)$\\
\hline
RTS&$1.113$&$1.122$&$ 1.014$&$ 1.014$&$1.060$&$1.060$&$ 1.048$&$ 1.051$\\
Eff&$0.927$&$0.874$&$ 1$&$ 0.999$&$1$&$0.999$&$ 0.943$&$ 0.915$\\
Num $Z$&--&--&$54$&$54$&$6$&$6$&$18$&$18$\\
\hline\hline
\end{tabular}
\end{center}
\end{table}

There are several interesting differences between Tables \ref{tab:SDF_prod_CD} and \ref{tab:SDF_prod_CD_bigZLessX}. First, in the ``No $Z$'' setting, we have substantially larger RTS, but fairly similar average technical efficiency. The results are naturally identical for the ``All $Z$'' setting. We also see that when the PSL is deployed that both land and labor are not selected and the inclusion of roughage serves to lower RTS on farm but we still see the evaporation of technical inefficiency. Interestingly, with more $Z$ to choose from the PSL selects fewer variables once land, labor and roughage are moved into the selection set (from 8 to 6). Finally, our PDL (both COLS and MLE) are consistent with the earlier results, but as with our PSL, we select fewer variables than when we had a smaller selection set (26 to 18). There is little difference in either RTS ($\approx$ 1.04 vs. $\approx$ 1.05) or average technical efficiency ($\approx$ [0.92-0.95] vs. $\approx$ [0.92-0.94]). We take comfort in the similarity of the PDL-COLS and PDL-MLE as it suggests a type of robustness to the size of the selection set. 

\subsection{Translog with All Inputs Mandatory}

While Table \ref{tab:SDF_prod_CD_bigZLessX} recognized that not all inputs may truly belong in the input set, we also believe that functional form mispecification could be an issue. Hence, in Table \ref{tab:SDF_prod_TL}, we repeat the analysis that produced Table \ref{tab:SDF_prod_CD}, but we use a translog functional form. The only change that this entails is that in the PDL (both COLS and MLE), instead of running five additional LASSOs (one for each input), we now run 20 additional LASSOs (one for each first-order term of input and for each second-order term of each input including interactions between any two inputs). 

\begin{table}[h]
\caption{Estimates of translog stochastic production frontier. Standard errors appear beneath each estimate in parentheses.}
 \label{tab:SDF_prod_TL} 
\begin{center}
\begin{tabular}{l|cccccccc}
\hline\hline
&\multicolumn{2}{c}{No $Z$}&\multicolumn{2}{c}{All $Z$}&\multicolumn{2}{c}{PSL}&\multicolumn{2}{c}{PDL}\\
&\multicolumn{1}{c}{COLS}&\multicolumn{1}{c}{MLE}&\multicolumn{1}{c}{COLS}&\multicolumn{1}{c}{MLE}&\multicolumn{1}{c}{COLS}&\multicolumn{1}{c}{MLE}&\multicolumn{1}{c}{COLS}&\multicolumn{1}{c}{MLE}\\
\hline
Feedstuffs&$ 0.341$&$ 0.319$&$ 0.457$&$ 0.457$&$ 0.416$&$ 0.409$&$ 0.355$&$ 0.342$\\
&$ (0.014)$&$ (0.014)$&$ (0.013)$&$ (0.013)$&$ (0.013)$&$ (0.014)$&$ (0.014)$&$ (0.014)$\\
Cows&$ 0.633$&$ 0.676$&$ 0.454$&$ 0.454$&$ 0.559$&$ 0.574$&$ 0.601$&$ 0.618$\\
&$ (0.024)$&$ (0.024)$&$ (0.021)$&$ (0.020)$&$ (0.020)$&$ (0.023)$&$ (0.023)$&$ (0.023)$\\
Land&$-0.011$&$-0.017$&$-0.017$&$-0.017$&$-0.010$&$-0.014$&$-0.012$&$-0.013$\\
&$ (0.010)$&$ (0.010)$&$ (0.009)$&$ (0.009)$&$ (0.009)$&$ (0.010)$&$ (0.010)$&$ (0.010)$\\
Labor&$ 0.021$&$ 0.014$&$-0.007$&$-0.007$&$-0.031$&$-0.033$&$ 0.005$&$ 0.000$\\
&$ (0.014)$&$ (0.013)$&$ (0.010)$&$ (0.010)$&$ (0.010)$&$ (0.012)$&$ (0.012)$&$ (0.012)$\\
Roughage&$ 0.093$&$ 0.088$&$ 0.122$&$ 0.122$&$ 0.129$&$ 0.126$&$ 0.079$&$ 0.079$\\
&$ (0.008)$&$ (0.008)$&$ (0.007)$&$ (0.007)$&$ (0.007)$&$ (0.007)$&$ (0.007)$&$ (0.007)$\\
\hline
RTS&$ 1.076$&$ 1.080$&$ 1.008$&$ 1.008$&$ 1.063$&$ 1.062$&$ 1.028$&$ 1.025$\\
Eff&$ 0.932$&$ 0.887$&$ 1$&$ 0.999$&$ 0.948$&$ 0.927$&$ 0.943$&$ 0.915$\\
Num $Z$&$ --$&$ --$&$51$&$51$&$23$&$23$&$46$&$46$\\
\hline\hline
\end{tabular}\end{center}
\end{table}

When we use all of the potential terms in our $Z$ matrix, we have a production process that produces the lowest RTS and the highest average efficiency score. For the COLS model the efficiency scores are all one as the corresponding residuals had positive skewness which precluded identification of $\sigma_u$ in a Normal-Half Normal specification of the composite error. MLE on the translog specification without $Z$'s has the lowest average efficiency (0.887) but also has a bizarre negative sign on the estimated input elasticity of land (this also happens across all of the models) which turns out not to be statistically significant. For our PDL results we see that all of the coefficients have positive signs except for Land with scale economies of 2.5\% (below the Cobb-Douglas PDL estimates in Table \ref{tab:SDF_prod_CD}) and an average farm efficiency score of 0.915 for PDL-MLE (compared to 0.951 in Table \ref{tab:SDF_prod_CD_bigZLessX} and 0.924 in Table \ref{tab:SDF_prod_CD}). For PDL-COLS this estimate is 0.943, same as in Table \ref{tab:SDF_prod_CD_bigZLessX} and slightly below that in Table \ref{tab:SDF_prod_CD} (0.954). Another interesting feature of Table \ref{tab:SDF_prod_TL} is that the PSL estimates (both COLS and MLE) do not suggest full efficiency (as in Tables \ref{tab:SDF_prod_CD} and \ref{tab:SDF_prod_CD_bigZLessX}) and their estimates of efficiency are similar to that of PDL, however some substantial differences remain for the estimates of input elasticities and RTS. Interestingly, note that out of the potential 51 $Z$'s, here PSL included 23, while PDL included 46.

\subsection{Translog with All Second Order Terms Optional}

Finally, we consider an alternative approach with the translog specification where we include all of the higher order terms of the translog, but treat them as variables that are selected via LASSO rather than making their inclusion mandatory. That is, we have a Cobb-Douglas specification and the quadratic and interaction terms for all the inputs (Land, Cows, Labor, Roughage and Feedstuffs) are part of $Z$ along with all of the other variables that we have included in there (for a total of 66 potential variables). This approach can be viewed as a way to optimally select a parsimonious model that improves upon the `too simplistic' Cobb-Douglas specification and `too big' translog specification. These estimates appear in Table \ref{tab:SDF_prod_CD_bigZwTL}.

\begin{table}[!h]
\caption{Estimates of Cobb Douglas stochastic production frontier using all $Z$ variables as well as second order terms from translog specification. Standard errors appear below in parentheses.}
 \label{tab:SDF_prod_CD_bigZwTL}
\begin{center}
\begin{tabular}{l|cccccccc}
\hline\hline
&\multicolumn{2}{c}{No $Z$}&\multicolumn{2}{c}{All $Z$}&\multicolumn{2}{c}{PSL}&\multicolumn{2}{c}{PDL}\\
&\multicolumn{1}{c}{COLS}&\multicolumn{1}{c}{MLE}&\multicolumn{1}{c}{COLS}&\multicolumn{1}{c}{MLE}&\multicolumn{1}{c}{COLS}&\multicolumn{1}{c}{MLE}&\multicolumn{1}{c}{COLS}&\multicolumn{1}{c}{MLE}\\
\hline
Feedstuffs&$ 0.386$&$ 0.360$&$ 0.457$&$ 0.457$&$ 0.439$&$ 0.439$&$ 0.343$&$ 0.320$\\
&$ (0.012)$&$ (0.013)$&$ (0.013)$&$ (0.013)$&$ (0.011)$&$ (0.013)$&$ (0.013)$&$ (0.013)$\\
Cows&$ 0.595$&$ 0.642$&$ 0.454$&$ 0.454$&$ 0.546$&$ 0.546$&$ 0.606$&$ 0.644$\\
&$ (0.020)$&$ (0.022)$&$ (0.021)$&$ (0.020)$&$ (0.017)$&$ (0.023)$&$ (0.022)$&$ (0.023)$\\
Land&$-0.010$&$-0.012$&$-0.017$&$-0.017$&$ 0.007$&$ 0.007$&$-0.012$&$-0.013$\\
&$ (0.009)$&$ (0.009)$&$ (0.009)$&$ (0.009)$&$ (0.008)$&$ (0.010)$&$ (0.011)$&$ (0.010)$\\
Labor&$ 0.035$&$ 0.032$&$-0.007$&$-0.007$&$-0.015$&$-0.015$&$ 0.001$&$-0.009$\\
&$ (0.012)$&$ (0.012)$&$ (0.010)$&$ (0.010)$&$ (0.009)$&$ (0.012)$&$ (0.012)$&$ (0.012)$\\
Roughage&$ 0.067$&$ 0.060$&$ 0.122$&$ 0.122$&$ 0.082$&$ 0.082$&$ 0.095$&$ 0.094$\\
&$ (0.005)$&$ (0.005)$&$ (0.007)$&$ (0.007)$&$ (0.004)$&$ (0.007)$&$ (0.007)$&$ (0.007)$\\
\hline
RTS&$ 1.074$&$ 1.082$&$ 1.008$&$ 1.008$&$ 1.059$&$ 1.059$&$ 1.034$&$ 1.036$\\
Eff&$ 0.930$&$ 0.892$&$ 1$&$ 0.999$&$ 1$&$ 0.999$&$ 0.940$&$ 0.903$\\
Num $Z$&$ --$&$ --$&$66$&$66$&$ 8$&$ 8$&$31$&$31$\\
\hline\hline
\end{tabular}\end{center}
\end{table}

In this case, the ``No $Z$'' estimates are identical to the ``No $Z$'' Cobb-Douglas estimates in Table \ref{tab:SDF_prod_CD}, while the ``All $Z$'' estimates are identical to the translog ``All $Z$'' estimates in Table \ref{tab:SDF_prod_TL}. The key differences lie with the PSL and PDL estimates. The PSL-COLS and PSL-MLE again produce estimates of efficiency that are 1 while the PDL-COLS and PDL-MLE estimates produce more reasonable estimates of average technical efficiency (0.94 and 0.9, respectively) and also select fewer variables than in the full PDL setting of Table \ref{tab:SDF_prod_TL} (31 out of 66 rather than 46 out of 51). We again also see that Land has a negative coefficient and is statistically insignificant across the four pairs of estimates. The coefficient on Labor appears to be highly influenced depending upon which set of $Z$ variables are included (both in terms of sign and significance). 

Comparing estimated elasticities of inputs from PDL-COLS with PDL-MLE, we see they are fairly similar, though with slightly bigger difference than in earlier cases, especially for Cows, which is obviously the most important input here. More importantly, we see that some of the estimates are substantially different from those estimated without $Z$ or with all of the $Z$'s for the PSL approaches. For example, the estimated coefficients for Feedstuffs is about 0.32-0.34 for PDL-MLE and PDL-COLS (with standard errors around 0.01), while they were 0.44 with all $Z$'s and 0.46 with PSL, i.e., well outside the two standard errors. This estimate is closer to that in COLS and MLE without $Z$'s (though still outside of two standard errors). The estimate of the coefficient for Cows in PDL is now much bigger than those from PSL and when all $Z$'s are included, while similar to that of MLE without any $Z$'s.  

Again, all these differences provide a vivid illustration of the sensitivity of the results  with respect to the choice of $Z$-variables, for both efficiency and RTS, as well as some key coefficients of frontier itself. Clearly, a practitioner in such situations usually (if not always) faces an important question from which policy conclusion may depend: Which of these results to trust? While we acknowledge that there is no one-size-fits-all answer for this question, as it depends on data and context, we would have higher confidence in the ``PDL-COLS/MLE'' results (especially MLE, yet with a double-check from COLS for robustness) due to the theoretical reasoning and the simulation evidence presented in this paper.

\section{Concluding Remarks}

We set out here to develop a coherent approach to applying machine learning methods to the realistic setting where practitioners are faced with a long list of potential technology shifters and what this means for estimating efficiency. We demonstrated that as the number of potential variables increases, this will mechanically lead to a finding of less (or no) inefficiency even if the variables included are irrelevant. While the LASSO and other high level machine learning methods can easily solve this problem, this introduces a selection bias that can corrupt estimation of other parts of the model. 

We documented that PDL, for either COLS or MLE, can effectively eliminate the selection bias issue, preserving both inefficiency and the structural interpretation of the production frontier. An application revealed several interesting differences between mechanical inclusion of a large set of potential technology shifters and the PSL and PDL approaches. This served as a reality test of our proposed model. 

While this paper adds to the nascent literature on machine learning methods applied to efficiency analysis, much remains to be done. Specifically, our focus here was on the inclusion of a large, and potentially irrelevant set of variables that capture technology differences, akin to the selective inattention interpretation of technology, but one could equally view a model where there exist a large number of determinants of inefficiency. This would entail non-trivial changes to the framework of our setup, possibly through nonlinear least squares or application of LASSO directly in the likelihood optimization. Another fruitful area would be to combine the two to determine how a variety of variables in both places of the model may impact selection and estimation.

Lastly, there are other important operational approaches to attend to. Our focus here is on the LASSO, but one could also think of application of generalized random forests or other machine learning methods, including ensemble methods. There exist a wide array of approaches for selection of the penalty parameter that may better align with the structure of the stochastic frontier model. Alternative sample splitting approaches may also prove interesting to explore. We end by drawing attention to recent work applying the LASSO to the dynamic panel GMM model of Arellano and Bond. This might be interesting for allowing dynamic efficiency. 



\clearpage


\bibliography{biblio}

\clearpage

\begin{appendix}
\section*{Appendix}
\subsection*{Neyman orthogonality for MLE: full set of moments}
The full set of moment conditions corresponding to the relevant first-order conditions of MLE derived by \citet[][Eqs. (11)-(13)]{ALS:77} is as follows: 
\[\begin{array}{cc}
\text{[A] }& \mathbb{E} x_i\left(\epsilon_i + r_i\right)=0 \\   
\text{[B] }& \mathbb{E} z_i\left(\epsilon_i + r_i\right)= 0 \\
\text{[C] }& \mathbb{E} \epsilon_ir_i= 0   
\end{array} \]
where $r_i=\phi\left(\rho\epsilon_i\right)/(1-\Phi\left(\rho\epsilon_i\right))$, $\rho=\lambda \sigma^{-1}$ and $\epsilon_i=y_i - \beta_0-x_i \beta - z_i\delta$. We omit the FOC for $\sigma^2$ because as shown in \citet[][]{ALS:77}, this can be concentrated out using the FOC for $\lambda$. What are the Neyman orthogonal moment conditions?

Similar to Section \ref{sec:NO}, a natural approach is to use in moment conditions [A]-[C] the parts of $y_i$ and $x_i$ orthogonal to $z_i$. Let $\epsilon^*_i := y^\perp_i - \beta_0-x^\perp_i\beta -z_i\delta$ and let $\phi^*_i:=\phi\left(\rho \epsilon^*_i\right)$ and $\Phi^*_i:=\Phi\left(\rho \epsilon^*_i\right)$. Then, we can define the inverse Mills ratio as $r^*_i:=\phi^*_i/(1-\Phi^*_i)$. Define what can be called conditional homoskewness, i.e., constant conditional third-order moment of $\epsilon_i|x_i, z_i$. Note that conditional homoskewness of $\epsilon_i|x_i, z_i$ implies conditional homoskewness of $\epsilon^*_i|x_i, z_i$ and of $r^*_i|x_i, z_i$. That is, $\mathbb{E}{\epsilon^*_i}^2r^*_i|x_i, z_i$, $ \mathbb{E}\epsilon^*_i{r^*_i}^2|x_i, z_i$, $ \mathbb{E}{r^*_i}^3|x_i, z_i$ and $\mathbb{E}{\epsilon^*_i}^3|x_i, z_i$ are scalars independent of $x_i, z_i$.  Denote 
$\mu_{21} = \mathbb{E}{\epsilon^*_i}^2r^*_i|x_i, z_i$ and $\mu_{12} = \mathbb{E}\epsilon^*_i{r^*_i}^2|x_i, z_i$. 

The moments $\mu_{21}$ and $\mu_{12}$ are not common and require some intuition. Since a Skew Normal random variable stems from truncation of a Bivariate Normal with correlation equal to $\rho$, the inverse Mills ratio captures the degree of truncation, and by default, the skewness of the resulting Skew Normal random variate. While the Skew Normal random variable that is at the heart of the stochastic frontier model under the Normal-Half Normal pair has constant skewness, the first order conditions from maximum likelihood do not reflect this skewness directly, but rather indirectly through the inverse Mills ratio. Why is this important in our context? $\mu_{21}$ can be viewed as a skewness adjusted variance and so we need this to be constant conditional on $x$ and $z$ otherwise, this would suggest that application of the double LASSO is changing the skewness, which impacts our FOCs and does not allow us to remove the nuisance parameters. The same is true for $\mu_{12}$ but in this case this FOC can be interpreted as a mean weighted moment where the weighting depends on the squared skewness. Again, if this were not constant then application of the double LASSO would impact the FOC in such a way that it becomes dependent upon the nuisance parameters. We do not view this as an overly restrictive assumption since we have already assumed that both $x$ and $z$ are independent of both $v$ and $u$, and hence $\epsilon$. 

Consider the moment conditions
\[\begin{array}{cc}
\text{[A*] }& \mathbb{E} x^\perp_i\left(\epsilon^*_i + r^*_i\right)=0 \\   
\text{[B*] }& \mathbb{E} z_i\left(\epsilon^*_i + r^*_i\right)= 0 \\
\text{[C*] }& \mathbb{E} \epsilon^*_ir^*_i= 0
\end{array} \]
These moment conditions correspond to running MLE where the dependent variable is the part of $y_i$ orthogonal to $z_i$ and the explanatory variables are $z_i$ and the part of $x_{i}$ orthogonal to $z_i$. 

To check when [C*] satisfies Neyman orthogonality we compute expected derivatives with respect to all the nuisance parameters: 
\[\begin{array}{cl}
\delta:& \mathbb{E}\left[ - z_i r_i^* \left( 1 - \rho^2 \epsilon_i^{*2} + \rho r_i^* \epsilon_i^* \right)\right] ,\\
\pi_x:& \mathbb{E} \left[- \beta z_i  r_i^* \left( 1 - \rho^2 \epsilon_i^{*2} + \rho r_i^* \epsilon_i^* \right) \right],\\
\pi_y:& \mathbb{E}\left[- z_i r_i^* \left( 1 - \rho^2 \epsilon_i^{*2} + \rho r_i^* \epsilon_i^* \right)\right]
.
\end{array} \] 

Outside of the multiplier $\beta$ on the $z_i$, these conditions are identical. 
Thus, Neyman orthogonality for the estimation of $\rho$ is achieved when
\begin{equation}\label{eq:no_for_rho}
  \mathbb{E} \big(z_ir_i^*\left(1-\rho
  \epsilon^*_i(\rho\epsilon^*_i-r^*_i)\right)\big)=0.
\end{equation}
This moment condition implies a restriction on the third-order moments of $(\epsilon_i^*, r_i^*)$, and specifically, on how they relate to $\mathbb{E} z_ir_i^*$. To see this, one can rewrite (\ref{eq:no_for_rho}) as follows
\begin{eqnarray}\nonumber
\mathbb{E} z_ir_i^* - \rho^2\mathbb{E} z_i{\epsilon_i^*}^2r_i^* + \rho\mathbb{E}z_i {\epsilon_i^*}{r_i^*}^2&=&\mathbb{E}z_ir_i^*-\rho^2\mathbb{E}_{x,z}z_i \mathbb{E}{\epsilon_i^*}^2r^*_i|x_i, z_i+ \rho\mathbb{E}_{x,z}z_i \mathbb{E}{\epsilon_i^*}{r^*_i}^2|x_i, z_i\\\nonumber
&=&\mathbb{E}z_ir_i^*-\rho^2\mu_{21}\mathbb{E}z_i + \rho\mu_{12}\mathbb{E}z_i \\\nonumber
  &=&\mathbb{E}z_ir_i^*+\rho (\mu_{12}-\rho\mu_{21}) \mathbb{E}z_i,
  \end{eqnarray}
where $\mu_{12}$ and $\mu_{21}$ are the constant conditional cross-moments of order three, defined above. 

If $z_i$ is demeaned ($\mathbb{E}z_i$=0) then (\ref{eq:no_for_rho}) becomes $\mathbb{E}z_ir_i^*=0$ which holds under independence of $z_i$ and $\epsilon_i^*$ and also implies [B$^*$]. Eq (\ref{eq:no_for_rho}) is also interesting as a component of a potential moment-based estimation of $\beta$ and $\rho$, which is robust to model misspecification.

\end{appendix}
\end{document}